\documentclass[aps,amsmath,amssymb,twocolumn]{revtex4-2}

\usepackage{graphicx}
\usepackage{indentfirst}
\usepackage{color,soul}
\usepackage{braket}
\usepackage{dsfont}

\begin{document}

\title{True image construction in quantum-secured single-pixel imaging\\under spoofing attack} 

\author{Jaesung Heo}
\author{Taek Jeong}
\author{Nam Hun Park}
\author{Yonggi Jo}
\email[]{yonggi@add.re.kr}
\affiliation{Advanced Defense Science \& Technology Research Institute, Agency for Defense Development, Daejeon 34186, Republic of Korea}

\date{\today}

\begin{abstract}
In this paper, we introduce a quantum-secured single-pixel imaging (QS-SPI) technique designed to withstand spoofing attacks, wherein adversaries attempt to deceive imaging systems with fake signals. Unlike previous quantum-secured protocols that impose a threshold error rate limiting their operation, even with the existence of true signals, our approach not only identifies spoofing attacks but also facilitates the reconstruction of a true image. Our method involves the analysis of a specific mode correlation of a photon-pair, which is independent of the mode used for image construction, to check security. Through this analysis, we can identify both the targeted image region by the attack and the type of spoofing attack, enabling reconstruction of the true image. A proof-of-principle demonstration employing polarization-correlation of a photon-pair is provided, showcasing successful image reconstruction even under the condition of spoofing signals 2000 times stronger than the true signals. We expect our approach to be applied to quantum-secured signal processing such as quantum target detection or ranging.
\end{abstract}

\maketitle

\section{Introduction}

Quantum security, security based on quantum phenomena, has been studied extensively in the quantum information field. Quantum key distribution (QKD) \cite{BB84,E91,BBM92} and blind quantum computation (BQC) \cite{Childs2005,Arrighi2006,Broadbent2009} are representative quantum information protocols that exploit quantum phenomena based quantum security including the no-cloning theorem \cite{Dieks1982,Wootters1982}, uncertainty relation \cite{Heisenberg1927,Busch2007}, and quantum measurement \cite{Cerf1998,Busch2003,Peres2004}. In both protocols, by analyzing changes of quantum states, or state errors, one can notice the existence of an adversary. Therefore, quantum security is fulfilled if information encoded in the quantum states is used only when there is no adversary.

In quantum sensing, there have been various studies on the rejection of environmental hindrances, especially external noise \cite{Lloyd2008,Liu2019,Kim2021APL,Blakey2022,Liu2023}. However, only a few studies have been conducted on preventing a spoofing attack, i.e., an attempt to deceive a sensing system by sending falsified signals to the system \cite{Malik2012,Roga2016,Yao2018,Heo2023}. A primary quantum-secured imaging (QSI) method was proposed in 2012 \cite{Malik2012} which provides threshold-type quantum security for an encoding mode, such as the polarization mode of a photon. Threshold-type quantum security means that there is a threshold error rate, which is an error rate obtained under the assumption of the optimal attack, and a protocol is interrupted when an error rate exceeds the threshold. Thus, in QSI, an obtained image is trusted only when a detected error rate is below the threshold; otherwise, it is discarded even if true information is included.

In QSI, encoding modes for a security check does not directly contribute to image formation, while QKD and BQC exploit encoding modes for both data processing and security analysis. This implies that QSI does not necessarily have the same form of security analysis as the previous quantum-secured protocols. Recently, there was a proposal for quantum-secured single-pixel imaging (QS-SPI) which tries to extract a true image under a certain kind of spoofing attack by an adversary \cite{Heo2023}. However, its security analysis considered only the optimal attack, which is an intercept-and-resend attack, so that it also provides threshold-type quantum security. Therefore, under a non-optimal attack, the extracted true image can be distorted.

In this paper, we present a QS-SPI method that can provide a true image under a spoofing attack. Our method exploits a mode-correlation of a photon-pair for security analysis. In our method, a type of spoofing attack is revealed by analyzing an erroneous image area and an error rate. With the detailed information of the attack, if true signals exist in detected signals, our method can reconstruct a true image even if the true signals are buried under strong fake signals. Thus, our method provides a new type of quantum security distinct from threshold-type security. We experimentally demonstrated our method with polarization-correlation of a photon-pair to compare reconstructed images of our method to a true image. We expect that adopting advanced techniques used in the existing quantum-secured protocols can further improve our method.

\section{Quantum-Secured Single-Pixel Imaging and spoofing attack}\label{Sec2}

\begin{figure}
	\includegraphics[width=0.45\textwidth]{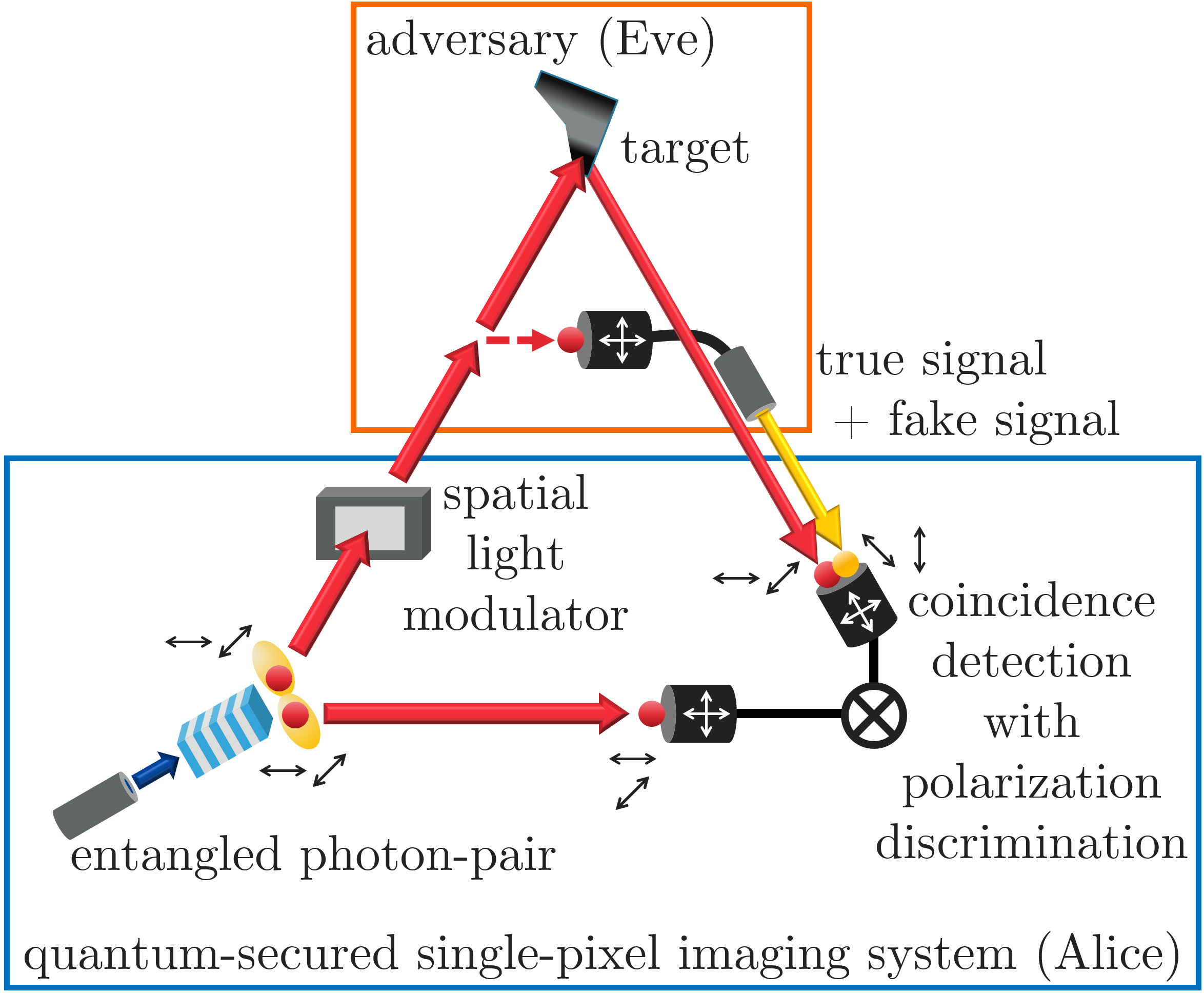}
	\caption{Conceptual scheme of QS-SPI under spoofing attacks. One photon of an entangled photon pair is sent to a target after spatial encoding with a spatial light modulator and is then measured. The other photon in the entangled pair is directly measured. The time-correlation and polarization-correlation of the two photons are analyzed from the measured data. An adversary can interact with all/partial signals and resend them to the detector for a spoofing attack.}
	\label{QSSPIconcept}
\end{figure}

The conceptual framework for QS-SPI is presented in Fig.~\ref{QSSPIconcept}. Initially, an entangled photon-pair is prepared. For illustration, we employ the polarization-correlation of the photon pair. However, alternative degrees-of-freedom may be exploited for QS-SPI. The signal photon of the photon-pair undergoes spatial intensity encoding through a spatial light modulator (SLM) before interaction with a target. After the interaction, measurements are conducted to obtain the polarization and timing information of the received photon. Simultaneously, the idler photon is directly measured to analyze the temporal and polarization correlations between two photons. As a result, spatial information of the target and polarization correct- and mismatched-coincidence rates of the photon pairs is obtained.

In the polarization measurement, one of two mutually unbiased bases (MUBs) is randomly chosen, such as a rectilinear basis and a diagonal basis. The rectilinear basis consists of horizontal and vertical polarization, and the diagonal basis consists of diagonal and anti-diagonal polarization. The polarization states are related with the following equations: $\ket{D}=(\ket{H}+\ket{V})/\sqrt{2}$ and $\ket{A}=(\ket{H}-\ket{V})/\sqrt{2}$, where $\ket{X}$ denotes a single photon $X$-polarization state, and $H$, $V$, $D$, and $A$ denote horizontal, vertical, diagonal, and anti-diagonal polarization, respectively.

In SPI, an image is constructed from the spatial correlation between encoding patterns on SLM and measured coincidence rates, i.e., $G(i,j)=\braket{P(i,j)I}-\braket{P(i,j)}\braket{I}$, where $G$ is a spatial correlation, $(i, j)$ is a pixel position, $P$ is ab intensity pattern, $I$ is a coincidence rate, and $\braket{\cdot}$ denotes the average for the whole $N$ trials \cite{Shapiro2008,Gibson2020}. Based on the correlation, image quality is influenced by the intensity pattern $P$ and the number of trials $N$. Therefore, using SPI with a larger number of diverse intensity patterns will yield a higher quality image.

During the target interaction phase, a user of the imaging system, called Alice, may encounter potential adversarial threats in the form of deceiving images by falsified data: a so-called spoofing attack. To accomplish the attack against SPI, an adversary, called Eve, should control Alice's coincidence rate according to Alice's spatial patterns. This can be realized by sending fake signals in diverse strategies such as substituting all/partial signals with fake signals, or illuminating strong patterned-jamming signals without blocking true signals to increase the accidental coincidence rates.

\section{Security analysis for spoofing attacks}\label{Sec3}

Due to the no-cloning theorem, Eve should interact with Alice's signal photon to obtain temporal and polarization information which introduces a change of the quantum state. Since we exploit two MUBs, if Eve's basis is different from Alice's, polarization of the prepared and received photons can be mismatched \cite{Malik2012,Heo2023}. To analyze the images constructed from the true and fake signals, the following spatial correlations are considered:
\begin{align}\label{QSSPI_img}
	\begin{split}
		G_\text{tot}(i,j) &= G_{T} (i,j) + G_{F} (i,j), \\
		G_\text{cor}(i,j) &= G_{T} (i,j) + (1-e_{F})G_{F}(i,j), \\
		G_\text{mis}(i,j) &= e_{F} G_{F}(i,j),
	\end{split}
\end{align}
where the subscriptions $T$ and $F$ denote true and fake, respectively. $G_{\text{tot}}$, $G_{\text{cor}}$, and $G_{\text{mis}}$ are obtained from all, correct, and mismatched polarization coincidence rates, respectively. These three spatial correlations are directly obtained from experimental data while true and fake spatial correlations are not. A fake signal error rate, $e_{F}$, is an error rate induced by Eve's signal only. Since it is determined by Eve's attack strategy, estimation of $e_{F}$ is the key to uncovering the attack strategy of an adversary; thus, a reconstructed image gets close to the true image with an exact estimation of $e_{F}$. Let us define a polarization state error rate, $e_{S}$, which is obtained from (mismatched coincidence rate)/(all coincidence rate). The relation between $e_{S}$ and $e_{F}$ becomes $e_{S}=e_{F} I_{F}/(I_{T}+I_{F})\leq e_{F}$. Therefore, $e_{S}=e_{F}\neq 0$ is satisfied only when there is no true signal. Note that $0.25 \leq e_{F}$ should be satisfied since 25\% is the fake signal error rate induced by the optimal attack: an intercept-and-resend attack \cite{Malik2012,Heo2023}.

To specify Eve's attack strategy, we define the following values:
\begin{align}\label{discrim_function}
	\begin{split}
		D_{1}(i,j)&:=\frac{G_{\text{mis}}(i,j)}{G_{\text{tot}}(i,j)}, \\
		D_{2}(i,j)&:=\frac{\left(G_{\text{mis}}(i,j)\right)^2}{e_S\left(G_{\text{tot}}(i,j)\right)^2}.
	\end{split}
\end{align}
If Eve interacts with all signals, i.e., there is no true signal, $E[D_{1}(i,j)]_{M}=e_{S}$ should be satisfied where an arithmetic mean of composite pixel values of $X(i,j)$ is written as $E[X(i,j)]$ and $M:=\{(i,j)|G_{\text{mis}}(i,j)\neq 0\}$ denotes an erroneous area. In this case, reconstruction of the true image is impossible. However, $E[D_{1}(i,j)]_{M}=e_{S}$ does not imply the absence of a true signal. Since the errorneous area $M$ and the fake signal error rate $e_{F}$ are independent variables, there may be instances where  $E[D_{1}(i,j)]_{M}=e_{S}$ is accidentally satisfied. To clarify Eve's strategy, $D_{2}$ should be exploited. Only when $E[D_{1}(i,j)]_{M}=E[D_{2}(i,j)]_{M}=e_{S}$ can we conclude that there are no true signals; otherwise, true image reconstruction is possible (See Appendix~\ref{PSA_discrimination}).

Image reconstruction is conducted with $e_{F}'$ which is the estimation of $e_{F}$. At the pixel $(i,j)$, $D_{1}=e_{F}$, if there is no true signal, $D_{1}<e_{F}$ when true signals exist. The condition $e_{S}\leq e_{F}\leq 1$ guarantees that $D_{1}\leq e_{S}$ implies the existence of true signals. Therefore, the estimation becomes more accurate for the region $M':=\{(i,j)|e_{S} < D_{1}(i,j) \leq 1\}$ rather than for $M$ since the pixels in $M'$ are closer to $e_{F}$. The estimated error rate is obtained from $e_{F}'=E[D_{1}(i,j)]_{M'}$. With $e_{F}'$ and Eq.~\ref{QSSPI_img}, a reconstructed true image $G_{T}'$ is formed based on the following equation:
\begin{align}\label{trustworthy}
	\begin{split}
		G_{T}'(i,j) &= G_\text{cor}(i,j) - \frac{1-e_{F}'}{e_{F}'} G_\text{mis}(i,j) \\
		&=G_{T}(i,j)-\delta G_{F}(i,j),
	\end{split}
\end{align}
where $\delta=e_F/e_F'-1$. If we set $e_{F}'=1/4$, the method becomes equivalent to the original QS-SPI when accounting for an intercept-and-resend type of spoofing attack \cite{Heo2023}.

Since no specific strategies for sending a fake signal are assumed, our method can be applied to any spoofing attack scenario. The true image reconstruction method is based on a QSI system that performs imaging and security analysis using the same data. If the two are executed with different data, $e_{S}$ has no relation with the images; thus, neither the attack discrimination nor reconstruction of the true image is possible.

In summary, security analysis for QS-SPI under a general spoofing attack is conducted as follows:
\begin{enumerate}
	\item Using $D_{1}$ and $D_{2}$, the possibility of true image reconstruction is determined as follows:
	\begin{itemize}
		\item $E[D_{1}(i,j)]_{M}\neq e_S$ : possible.
		\item $E[D_1(i,j)]_{M}=e_S$ \& $E[D_2(i,j)]_{M} > e_S$ : possible.
		\item $E[D_1(i,j)]_{M}=E[D_2(i,j)]_{M}=e_S$ : impossible.
	\end{itemize}
	\item If the reconstruction is possible, Alice can obtain a credible image by Eq.~\ref{trustworthy} with $e_{F}'=E[D_{1}(i,j)]_{M'}$.
\end{enumerate}

\section{Proof-of-Principle Demonstration}\label{Sec4}

\begin{figure*}[t!]
	\includegraphics[width=0.95\textwidth]{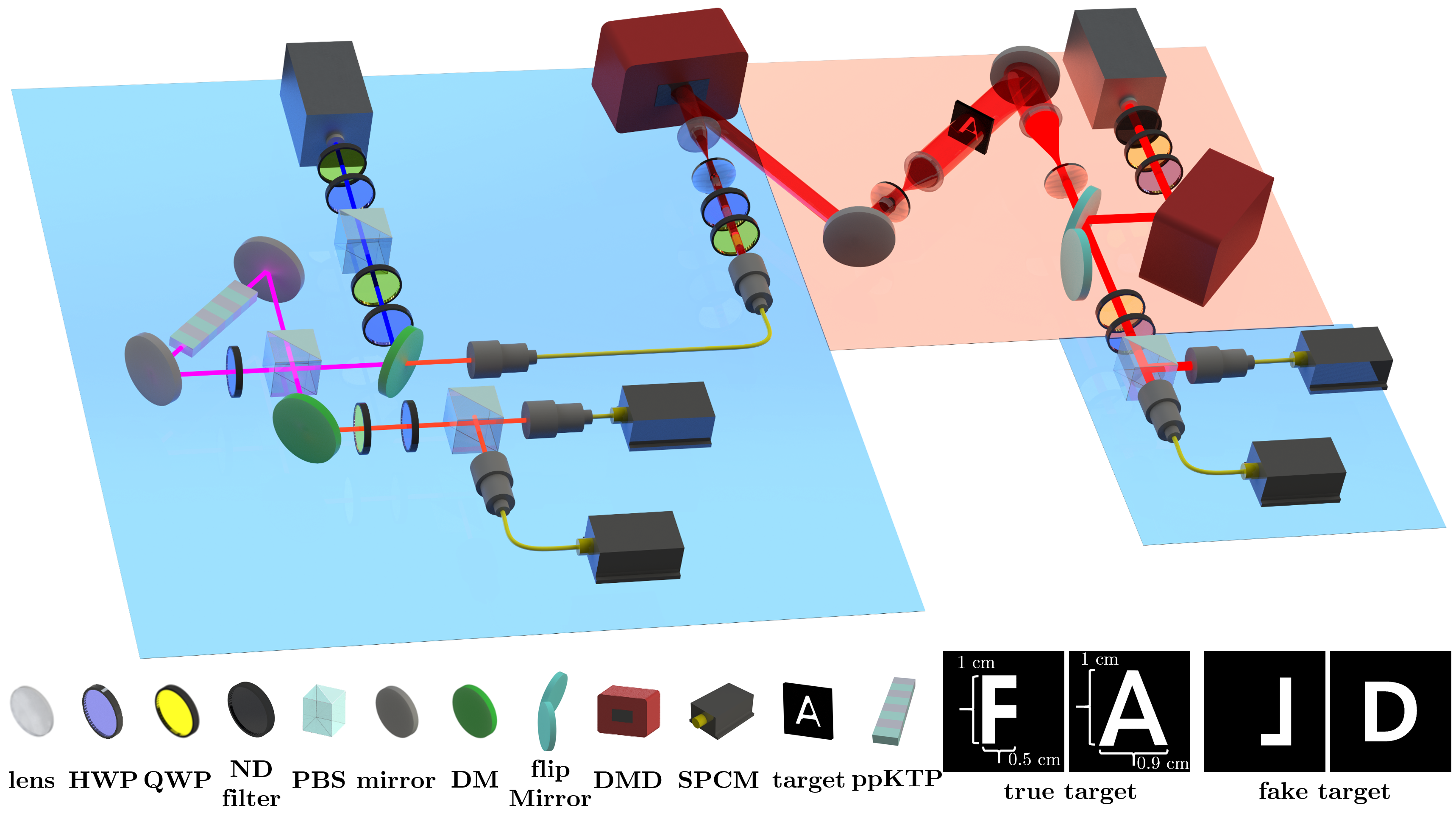}%
	\caption{Schematic of the experimental setup. The blue and red regions denote Alice's setup and the exterior including Eve's setup, respectively. In the blue region, polarization-entangled photon pairs are generated by a ppKTP crystal inside the Sagnac interferometer structure, and a single photon is detected by single photon counting modules with polarization discrimination. Eve sends a fake signal with fake target information encoded by her DMD. With a flip mirror, either true or fake signals are selected for detection. HWP: half-wave plate; QWP: quarter-wave plate; ND: neutral density; PBS: polarizing beam splitter; DM: dichroic mirror; SPCM: single photon counting module; ppKTP: periodically poled potassium titanyl phosphate.}\label{exp}
\end{figure*}

Fig. \ref{exp} shows a schematic of the QS-SPI experimental setup. In Alice's setup, depicted by the blue region in the figure, 810 nm entangled photon pairs were created by pumping a 10 mm-long periodically poled potassium titanyl phosphate (ppKTP) crystal (Raicol Crystals) using a 405 nm continuous wave (CW) laser (Toptica, TopMode). The ppKTP is located inside a Sagnac interferometer setup for generating the polarization entangled Bell state \cite{Kim2006}, $\ket{\Phi^{+}}=\left(\ket{H,H}+\ket{V,V}\right)/\sqrt{2}$. The fidelity of the generated state was 98.6\%. The entangled pairs are detected by single photon counting modules (SPCMs, Excelitas Technologies, SPCM-780-13-FC) with polarization discrimination. The bases are randomly chosen by wave plates, but the bases of the signal and idler are always identical. A digital micromirror device (DMD, Vialux GmbH, DLP650LNIR) was exploited for spatial intensity modulation. For image construction in SPI, the coincidence rate of the signal and idler SPCMs is utilized rather than the single photon rate of each mode. In the setup, the pump power was 5 mW, with signal and idler photon rates at the SPCMs without a target being $6\times 10^{3}$ and $8\times 10^{4}$, respectively. The coincidence rate of the same polarization without a target was approximately 300 cps. We used a specific set of orthogonal patterns known as Hadamard patterns \cite{Pratt1969, Souza1988}, with a resolution of $32\times 32$. In our demonstration, each pattern required two shots to represent the $-1$ element in the Hadamard patterns. Since there are 1024 Hadamard patterns at $32\times 32$ resolution, a complete image was constructed with 2048 shots. The coincidence window was set to 650 ps, and the photon acquisition time for each shot was 3.5 s.

\begin{figure*}[t!]
	\includegraphics[width=0.95\textwidth]{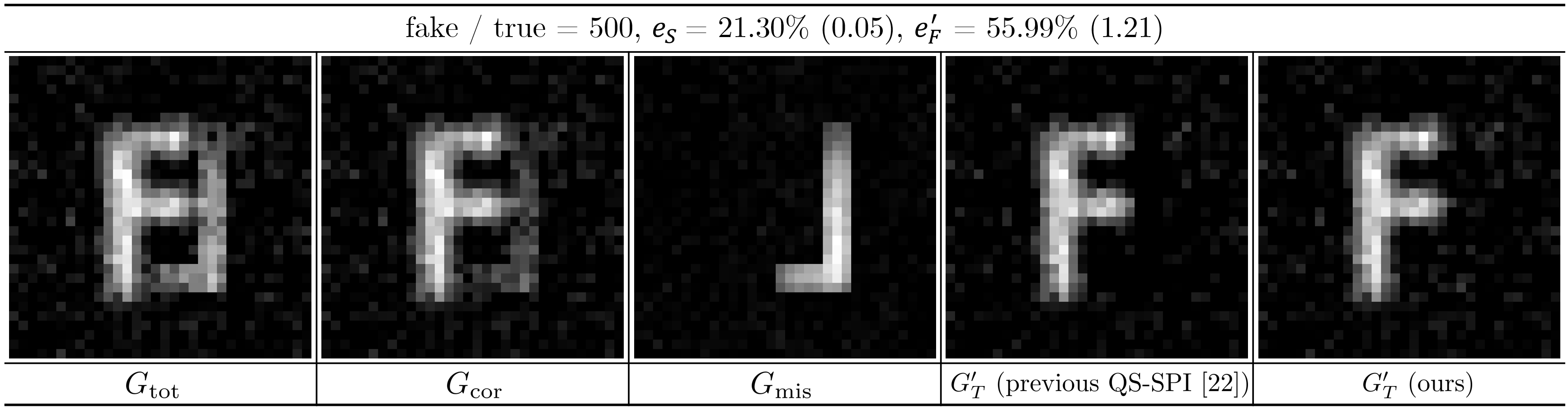}%
	\caption{Obtained images under a patterned-jamming attack with random polarization ($e_{F} = 50\%$). The false image obtained by SPI is the digital number ``8", but the true image is ``F". Both the previous QS-SPI \cite{Heo2023} and ours can reconstruct ``F"; however, the previous QS-SPI overly deletes the $G_{\text{mis}}$ area. This over-deletion can distort the true image, as will be demonstrated in the next experimental results. The estimated error rate is $e_{F}' = 55.99\%$ and its standard deviation is $(1.21)$.}\label{PSA_F8_exp}
\end{figure*}

Note that our SPI setup is designed as a proof-of-principle demonstration for our security analysis and true image reconstruction method. Therefore, advanced techniques can be applied for various purposes. For instance, while the pattern set in SPI affects image quality and acquisition time, quantum security is independent of the pattern. Consequently, other sets, such as Fourier \cite{Zhang2017} or discrete cosine transform \cite{Liu2017} patterns, can be used to enhance image quality and reduce the sampling ratio. Furthermore, cutting-edge devices can also enhance the performance of our setup. For example, narrowing the coincidence window using SPCMs and TCSPC with smaller electronic timing jitter would enable the system to better reject stronger light from Eve.

Eve's setup, shown in red in Fig. \ref{exp}, is composed of an 810 nm CW laser (homemade external-cavity diode laser, Thorlabs, M9-808-0150) and another DMD. The power of Eve's laser required to cause accidental coincidences can be calculated as follows. If the idler photon rate is $N_{I}$, Eve's photon rate is $N_{F}$, and the coincidence window is $\tau$, then the accidental coincidence rate is $\tau N_{I} N_{F}$. This rate matches the coincidence rate of the entangled-photon pair source, 300 cps, when $N_{F}\sim 5.8\times 10^{6}$, which is 1000 times larger than the original signal photon rate. For an 810 nm CW laser, this corresponds to a power of approximately 1.41 pW for Eve's photon rate. Eve's DMD encodes fake target information to fake signals by displaying an overlap of an Alice's imaging pattern and a fake target image. We demonstrated two targets: a true target ``A'' having overlap with a fake target ``D,'' and a true target ``F'' having no overlap with a fake target. Both are combined to show the form of the digital number ``8.''

\begin{figure*}[t!]
	\includegraphics[width=0.95\textwidth]{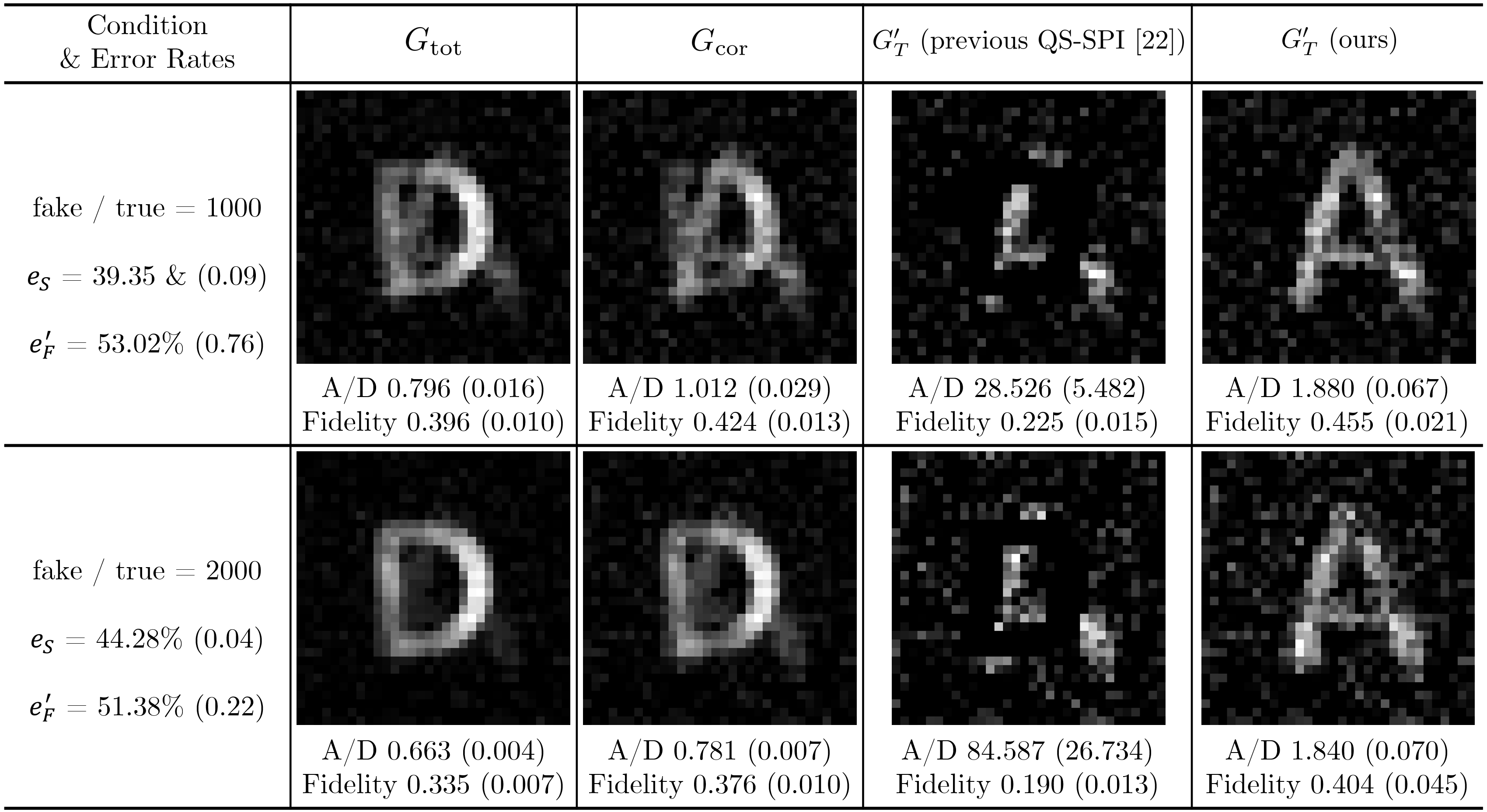}
	\caption{Images obtained through the QS-SPI system under a patterned-jamming attack with random polarization ($e_{F} = 50\%$). The fake signal is 1000 and 2000 times stronger than the true signal. The fake image is the letter ``D," while the true image is ``A." The true image is well reconstructed by our QS-SPI, while the fake signal area is overly deleted with the previous QS-SPI \cite{Heo2023}. Image qualities were compared by using the A to D ratio (A/D) and the fidelity to the ideal image ``A," each denoted with standard deviation in parentheses.}\label{PSA_RP_exp}
\end{figure*}

By controlling the flip mirror, true and fake signals are selectively received. When the flip mirror blocked the true signal and reflected the fake signal, only the fake signal was detected, and if the flip mirror did nothing to the signals, the true signal was detected. Since we do not have an on-demand single-photon generator, an attack was simulated by blocking a true signal and illuminating strong light for accidental coincidence. Eve's attack was simulated by mixing the two data. The detections were analyzed by coincidence counts in each polarization combination. Two attack strategies are demonstrated: a patterned-jamming attack with random polarization and an intercept-and-resend attack. The details of the demonstration are given in Appendix~\ref{exp_details}.

Fig.~\ref{PSA_F8_exp} shows the obtained images under a patterned-jamming attack with random polarization when the fake signal is 500 times stronger than the actual signal. A target is the letter ``F," and this attack makes the imaging system construct the digital number ``8" by using fake signals. All images shown are normalized to a scale of 0 to 255 pixel values. The theoretical fake signal error rate is $e_{F}=50\%$ for this attack, i.e., if there is only fake signal, the error rate becomes $50\%$. The polarization state error rate of the obtained image is $21.30\%$ with a standard deviation of $0.05$, making it undetectable by the original QSI \cite{Malik2012}. The digital number ``8" is constructed from both the original SPI ($G_{\text{tot}}$) and polarization-filtered SPI ($G_{\text{cor}})$. The previous QS-SPI \cite{Heo2023} and our QS-SPI can construct the true image ``F". The previous QS-SPI overly deletes the area of $G_{\text{mis}}$, making the background area appear clearer than in our method. However, this over-deletion can distort the true image, as will be demonstrated in the following results. The estimated fake signal error rate is $55.99\%$ $(1.21)$.

\begin{figure*}[t!]
	\includegraphics[width=0.95\textwidth]{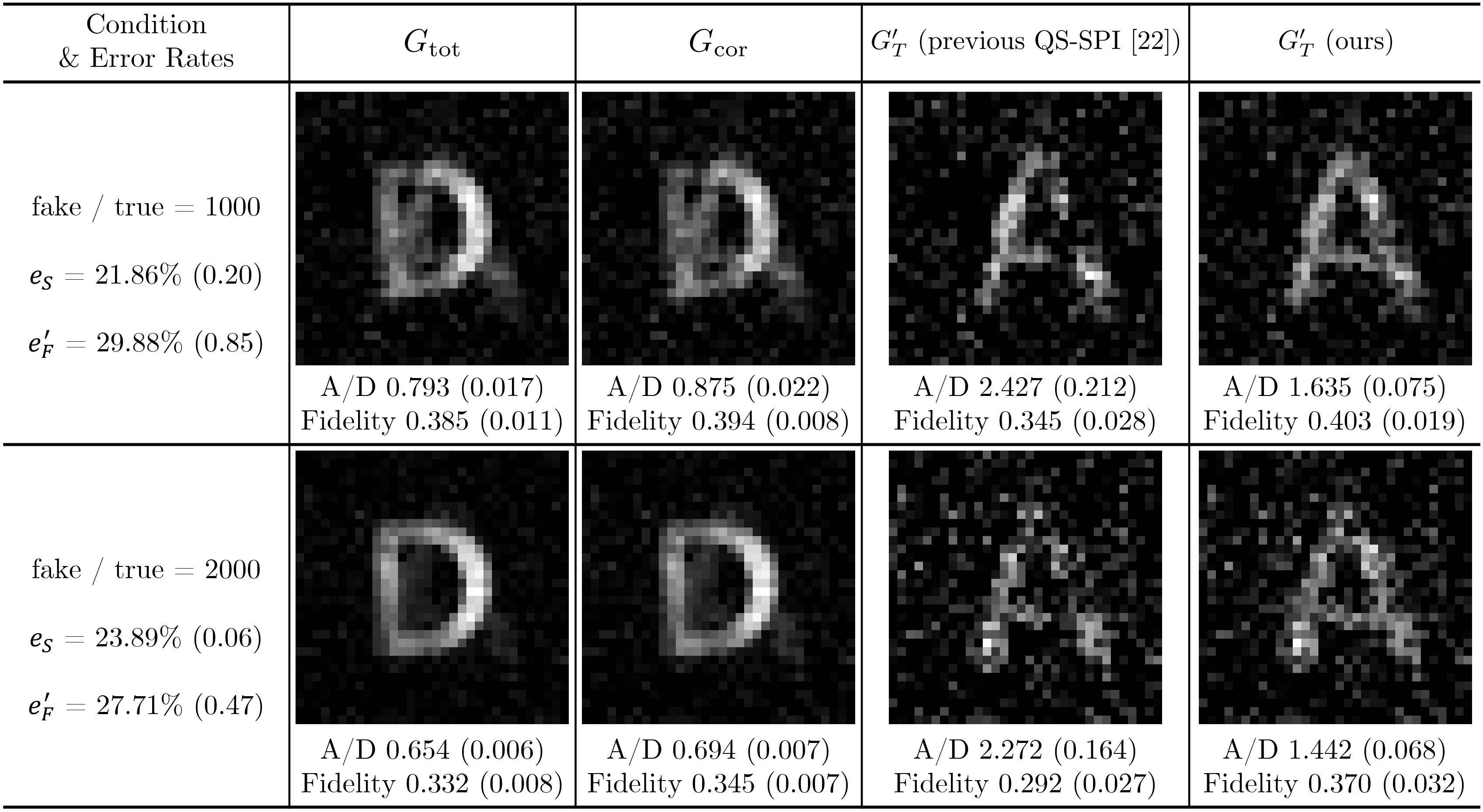}%
	\caption{Images obtained through the QS-SPI system under an intercept-and-resend attack ($e_{F} = 25\%$). The fake signal is 1000 and 2000 times stronger than the true signal. The fake image is the letter ``D," while the true image is ``A." The true image is well reconstructed by our QS-SPI and the previous QS-SPI \cite{Heo2023}; however, ours has better image quality. Image qualities were compared by using the A to D ratio (A/D) and the fidelity to the ideal image ``A" with their standard deviations.}\label{PSA_ira_exp}
\end{figure*}

Fig.~\ref{PSA_RP_exp} shows the obtained images under a patterned-jamming attack with random polarization when the true target is the letter ``A," and the fake target is ``D." The fake signal is 1000 and 2000 times stronger than the true signal. Similar to the previous image, the original SPI and SPI with polarization filtering cannot obtain the true image, while our QS-SPI can reconstruct a true image. Since this is not an optimal attack, the previous QS-SPI \cite{Heo2023} overly deletes the fake image; thus, the reconstructed image is far from the true image. This implies that Eve's attack is successful if her goal is to ruin the true image rather than display a fake one, as the previous QS-SPI can detect the attempt but fails to reconstruct the true image.

To compare the quality of the true image, we exploit two measures: the A to D ratio and fidelity. First, the fidelity is obtained with the true target ``A" shown in Fig.~\ref{exp}. Let us define the spatial correlation of the true target as $G_{A}(i,j)$, then $G_{A}(i,j)=1$ for the ``A" region and $0$ otherwise. The fidelity is calculated from $E[G_{A}(i,j)G_{\text{exp}}(i,j)]$ where $G_{\text{exp}}$ is an experimentally constructed spatial correlation. A correlation of the fake target $G_{D}(i,j)$ can be obtained in the same way. We can then define the A to D ratio as $E[G_{A}(i,j)G_{\text{exp}}(i,j)]/E[G_{D}(i,j)G_{\text{exp}}(i,j)]$. The A to D ratio quantifies the restoration of the true image and rejection of the fake signal. The fidelity is used to quantify the quality of the true image. Here, we compared the A to D ratio of $G_{\text{tot}}$, $G_{\text{cor}}$, and $G_{T}'$ (ours), and the fidelity of $G_{T}'$ of the previous QS-SPI and ours.

From the A to D ratio, the reconstructed image obtained by our QS-SPI is closer to the true image and farther from the fake image compared to $G_{\text{tot}}$ and $G_{\text{cor}}$. The image obtained by the previous QS-SPI has the highest A to D ratio since the pixels in the fake image region are mostly 0 due to over-deletion. The fidelity shows that the image obtained by our QS-SPI is closer to the true image than that by the previous QS-SPI.

In principle, our true image reconstruction method operates independently of the ratio between fake and true signals. Due to the saturation of our SPCMs, we were unable to demonstrate QS-SPI under stronger fake signals. Nevertheless, the results effectively showcase the capability of our method, as it successfully reconstructs the true image even when the fake image dominates in $G_{\text{tot}}$.

Fig.~\ref{PSA_ira_exp} shows obtained images under an intercept-and-resend attack. Since the state error rate is below $25\%$, the original QSI cannot detect this attack. Different from the patterned-jamming attack case, the previous QS-SPI \cite{Heo2023} well reconstructs the true image. Comparing the qualities of the images obtained by the two QS-SPI protocols, the previous QS-SPI is better than ours in the A to D ratio. However, as shown in Fig.~\ref{PSA_RP_exp}, the previous QS-SPI overly deletes the fake image information since the protocol works under the assumption of an ideal intercept-and-resend attack. From our QS-SPI, the estimated fake signal error rates, $29.88\%$ and $27.71\%$, are close to the ideal fake signal error rate of an intercept-and-resend attack: $25\%$. However, the existence of a small variation means that the intercept-and-resend attack is not perfectly demonstrated for some experimental reasons. Thus, the previous QS-SPI still overly deletes the fake signal information in the demonstration. Therefore, our QS-SPI provides higher fidelity compared to the previous QS-SPI.

\begin{figure}[t!]
	\includegraphics[width=0.49\textwidth]{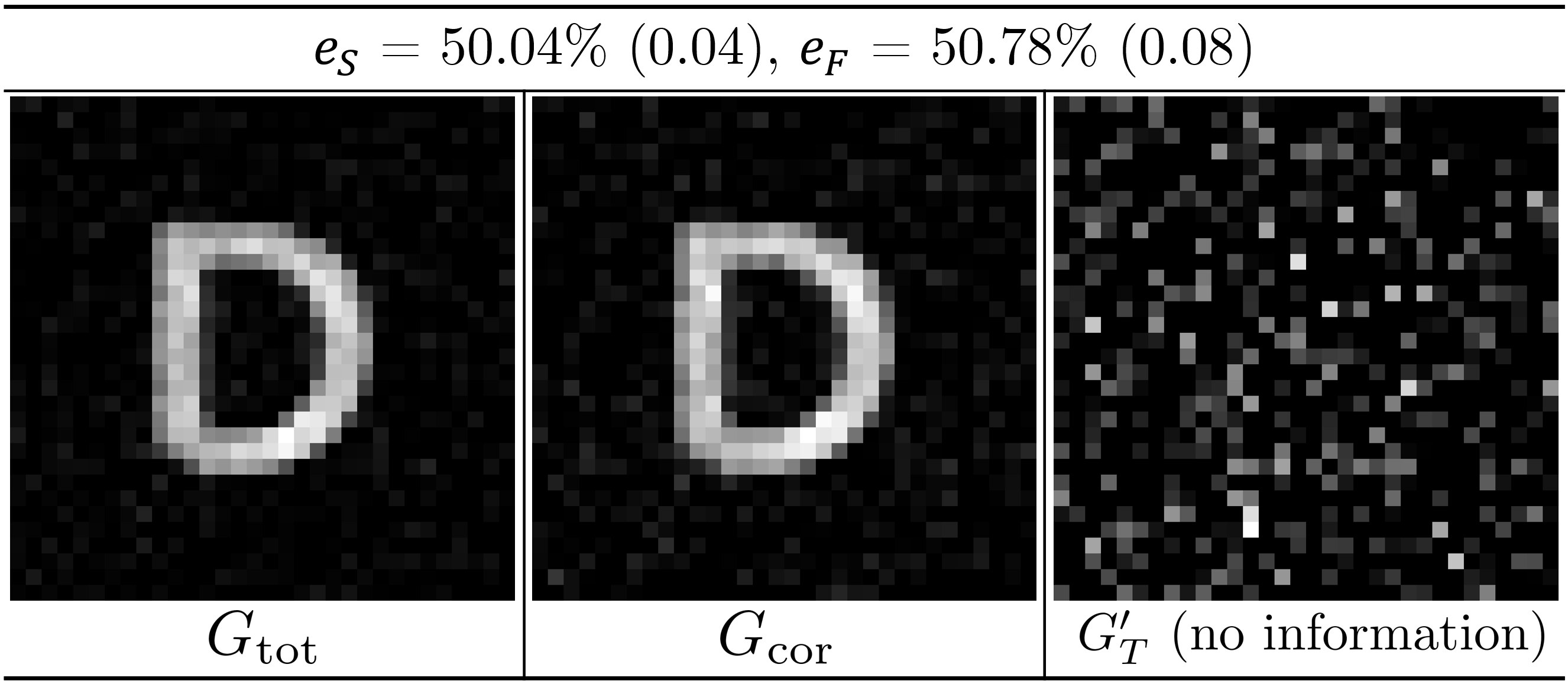}%
	\caption{Images obtained through the QS-SPI system under a patterned-jamming attack with random polarization and no true signals. As $e_S$ and $e_F$ are very similar with less than a 2\% difference, we can conclude that there are no true signals. This is supported by the result of $G_{T}'$ showing no image information.}\label{FSA_exp}%
\end{figure}

Lastly, Fig.~\ref{FSA_exp} shows the obtained images under a patterned-jamming attack with random polarization and no true signals. The state error rate is $50.04\%$ and the fake signal error rate is very similar at $50.78\%$. Thus, we can conclude there is no true signal. This is also verified from the image reconstruction $G_{T}'$; since there is no true signal, only white noise is shown in the reconstructed image.

\section{Summary and Discussion}\label{Sec5}

In this paper, the true image reconstruction method in QS-SPI under a general spoofing attack is presented. Like QKD, the QSI \cite{Malik2012} and the original QS-SPI \cite{Heo2023} considered only the optimal attack and provided threshold-type quantum security, i.e., if the error rate is higher than the error threshold, the protocol is interrupted even when true signals exist. Thus, the previous protocols can identify but not prevent Eve's effort to distort the true image rather than present a fake one, such as through intercept-and-resend attacks using entirely different polarization. However, our method does not assume a specific attack strategy, and thus, the type of spoofing attack can be discovered by using our method. Moreover, it is possible to reconstruct a true image under various types of spoofing attacks. A proof-of-principle demonstration of our method is provided, and we show the reconstructed true images with our method have better quality compared to the original one. Because our experimental setup shares similarities with heralded SPI \cite{Kim2021APL}, we anticipate that our method is also resilient to jamming attempts using strong chaotic light.

In our method, the fake error rate $e_{F}$ is estimated from areas where most of the signals are fake. If Eve's attack area completely overlaps with a target area, it becomes challenging for the estimated fake error rate $e_{F}'$ to match $e_{F}$ accurately in such scenarios. However, the gap between $e_{F}$ and $e_{F}'$ increases when the number of fake signals is fewer than that of true signals. This means that the impact of fake signals on the image is small when this gap is large. Therefore, constructing the exact true image under a fully overlapped attack is difficult, but the resulting image may still closely resemble the true one.

Note that the security framework offered by QS-SPI differs from that of classical secure SPI studies. Classical secure SPI research primarily aims to prevent a third party from obtaining the same image as the authorized party \cite{Zhang2018, Ye2019}. This is typically achieved by masking intensity patterns on the SLM with encrypted patterns, ensuring only those with the correct decryption keys can reconstruct the image. In contrast, QS-SPI focuses on a different aspect of security. It aims to prevent a third party from manipulating the imaging system to display a fake image by exploiting quantum phenomena.

Similar to the previous QS-SPI \cite{Heo2023}, we expect that the security of our QS-SPI can be enhanced with the advanced techniques exploited in quantum secure communication, such as protocols based on three mutually unbiased bases \cite{Bruss1998}, high-dimensional quantum states \cite{Cerf2002, Jo2016, Bouchard2018, Jo2019}, multipartite entangled state \cite{Chen2007,Proietti2021}, and hyper-entangled states \cite{Wang2015, JKim2021}. Although the demonstration of our method is limited to an SPI system in this paper, the methods are expected to be adopted in other applications such as quantum target detection \cite{Lloyd2008, Lee2023} or quantum target ranging \cite{Zhuang2021,Qian2023}. In particular, LiDAR \cite{Kim2021,Reichert2022,Lim2023,Lee2023POSTECH,Reichert2023} can provide higher security not only against a jamming attack with external noise rejection \cite{Liu2019,Kim2021APL,Blakey2022,Liu2023}, but also against a spoofing attack with our method.

The security of QS-SPI based on the quantum physics, while the true image reconstruction method is facilitated through data processing. Our experimental setup has not allowed us to verify whether our data processing functions effectively under external chaotic light, as SPI with heralded photons can reject strong chaotic light \cite{Kim2021APL}. We anticipate that our data processing could also remove portions of an image affected by external intense light sources. If successful, this approach could be adapted for noise reduction in diverse active imaging systems, encompassing not only SPI but also multi-pixel imaging technologies.

\acknowledgments{This work was supported by the Agency for Defense Development Grant funded by the Korean Government.}

\section*{Appendice}
\appendix
	\section{Spoofing attack with true and fake signal mixing}\label{PSA_discrimination}
	From Eq.~\ref{QSSPI_img} and Eq.~\ref{discrim_function}, $D_1(i,j)$ under a spoofing attack by mixing true and fake signals is 
	\begin{align}\label{d1_PSA}
		D_{1}(i,j)=
		\begin{cases}
			e_{F} & G_{T}(i,j)=0 \\
			e_{F}\frac{G_{F}(i,j)}{G_{T}(i,j)+G_{F}(i,j)} & G_{T}(i,j)\neq0
		\end{cases}.
	\end{align}
	This shows that the presence of $G_{T}(i,j)$ lowers pixel values below $e_{F}$. Therefore, if $E[D_{1}(i,j)]\neq e_{S}$, a true signal exists. However, $E[D_{1}(i,j)]= e_S$ does not guarantee that there is no true signal. For example, if $G_{T}$ exists in half of $M$ and $G_{T}(i,j)=G_{F}(i,j)$ inside $M$ and $e_{S}=0.75 e_{F}$, then $E[D_{1}(i,j)]=e_{S}$, although  a true signal exists. Still, the cases are distinguishable by using $D_{2}(i,j)$.
	
	Failure to discriminate the two cases is equivalent to the following: with true signals, if $E[D_{1}(i,j)]=e_{S}$, then $E[D_{2}(i,j)] = e_{S}$. To analyze this statement, let us define the $l$-th area in $M$ as $A_{l}$ where $G_{T}(i,j)\neq 0$ and the composite pixel values are identical to a constant value $v_{l}$. Therefore, in that area, $\frac{G_{F}(i,j)}{G_{T}(i,j)+G_{F}(i,j)} = v_l$. Assume a total of $q$ areas exist in $M$ where $G_{T}(i,j)\neq 0$. Lastly, let us define the area where $G_{T}(i,j)=0$ inside $M$ as $A_{q+1}$. Since we assume that there is a true signal and $E[D_{1}(i,j)]=e_{S}$, $G_{T}(i,j)\neq 0$ exists in $M$, then, $E[D_{1}(i,j)]=e_{S}$ gives
	\begin{align}\label{d1_calculation}
		e_{F}\frac{A_1v_1+A_2v_2+\dots+A_qv_q+A_{q+1}}{A_1+A_2+\dots+A_q+A_{q+1}}=e_{S}.
	\end{align}
	Using Eq.~\ref{d1_calculation}, $E[D_{2}(i,j)]/e_{S}$ is
	\begin{widetext}
	\begin{align}
		\frac{E[D_{2}(i,j)]}{e_{S}}&=
		\left(\frac{e_{F}}{e_{S}}\right)^2\frac{A_1(v_1)^2+A_2(v_2)^2+\dots+A_q(v_q)^2+A_{q+1}}{A_1+A_2+\dots+A_q+A_{q+1}} \\
		&=\frac{\left(A_1+\dots+A_q+A_{q+1}\right)
			\left(A_1(v_1)^2+\dots+A_q(v_q)^2+A_{q+1}\right)
		}{
			\left(A_1v_1+A_2v_2+\dots+A_qv_q+A_{q+1}\right)^2
		} \geq 1,
	\end{align}
\end{widetext}
	where the last inequality is the Cauchy-Schwarz inequality. For $E[D_{2}(i,j)]=e_{S}$ to be accomplished, the last equality should be satisfied. However, equality is reached only when all $v$ values are equal to 1, indicating that $G_{T}(i,j)=0$ in $M$. This contradicts the assumption that $E[D_{1}(i,j)]=e_{S}$. Therefore $D_{1}$ and $D_{2}$ together can always discriminate the existence of a true signal. If $E[D_{1}(i,j)]=e_{S}$ and $E[D_{2}(i,j)]>e_{S}$, then there is a true signal; if $E[D_{1}(i,j)]=E[D_{2}(i,j)]=e_{S}\geq 25\%$, then there is no true signal.

	\section{Demonstration details}\label{exp_details}
	
	Spatial patterns are crucial for the quality of SPI \cite{Nie2021,Zhang2022,LIN2023}. We exploited a specific orthogonal pattern set known as Hadamard patterns \cite{Pratt1969, Souza1988, Duarte2008, Kim2021APL, Heo2022}, but alternative sets can also enhance SPI image quality with reduced sampling ratios, such as Fourier or discrete cosine transform patterns \cite{Zhang2017, Liu2017}. Note that spatial patterns influence SPI image quality but do not impact our quantum security.
	
	The Hadamard matrix of order $2^{n+1}$ is constructed as
	\begin{align}
		H_{2^{n+1}}=H_{2^{n}}\otimes H_{2},
	\end{align}
	where
	\begin{align}
		H_{2}=\begin{pmatrix}
			1 & 1 \\
			1 & -1
		\end{pmatrix},
	\end{align}
	and $\otimes$ denotes the tensor product. Reshaping each row of a Hadamard matrix of order $2^{2n}$ into a square matrix, a total of $2^{2n}$ Hadamard patterns of $2^{n}\times 2^{n}$ resolution are obtained. The patterns include negative elements. To display the pattern set to an SLM with intensity modulation such as a digital-micromirror-device (DMD), two shots are required for a single pattern: a pattern formed by transitioning $+1$ elements as white and $-1$ as black, and the opposite \cite{Gibson2020}. Therefore, a total of $2^{2n+1}$ shots are required for a $2^n \times 2^n$ resolution image. This can be reduced to half by one-shot detection of a single pattern through various techniques \cite{Yu2021}.
	
	We used Hadamard patterns reconstructed based on a Hadamard matrix of order $2^{10}$. We used all 2048 shots for image construction, but as demonstrated in the heralded SPI experiment \cite{Kim2021APL}, it's also possible to use only 700 shots out of the 2048. However, reducing the sampling ratio will lead to degraded image quality. The resolution of the images is $32\times32$.
	
	To simulate a spoofing attack, fake signals are illuminated to the detection system for an accidental coincidence. Due to the loss of true signals in the target interaction, fake signals can induce accidental coincidence. By controlling the power of the illumination, Eve can manipulate Alice's coincidence rates. The fake signal is sent in either the $H$- or the $D$-polarization, randomly; therefore, the raw data represents a patterned-jamming attack with random polarization. The intercept-and-resend attack is demonstrated as follows. First, as the error rate in the mismatched bases selection of Alice and Eve is also $50\%$ in this attack, the raw data is used without modification. If Alice and Eve choose identical bases, ideally no error is induced, so only the correct data should be exploited for image construction. Therefore, the error coincidences should be discarded, meaning that we discard one out of the two possibilities of Eve's successful attack. To account for this, we double the coincidence counts of the selected data. In total, $e_{F} = 25\%$ is made \cite{Heo2023}.

\end{document}